\documentclass[aps,pre,floats,twocolumn,twoside,floatfix,pacs]{revtex4}
\usepackage{epsfig}
\usepackage{amsmath}
\usepackage{amssymb}
\newcommand{\mus}{\mu_{\scriptstyle s}}

\begin{document}

\title{Fluctuation relation and heterogeneous superdiffusion in glassy
  transport}
            
\author{Mauro Sellitto} \affiliation{Laboratoire de Physique et
  M\'ecanique des Milieux H\'et\'erog\`enes \\ ESPCI - 10, rue
  Vauquelin - 75231 Paris Cedex 5, France}

\begin{abstract}
  Current fluctuations and related steady state fluctuation relation
  are investigated in simple coarse-grained lattice-gas analogs of a
  non-Newtonian fluid driven by a constant and uniform force field, in
  two regimes of small entropy production.  Non-Gaussian current
  fluctuations and deviations from fluctuation relation are observed
  and related to the existence of growing amorphous correlations and
  heterogeneous anomalous diffusion regimes.
\end{abstract}

\maketitle

Fluctuation theorem (FT) and nonequilibrium work relations are results
of remarkable generality representing an important step towards the
formulation of statistical mechanics far from
equilibrium~\cite{HaSc,Ri,Vulpio}.  FT states that the ratio of
probabilities of observing an entropy production $W_{\tau}$ over a
long time interval $\tau$, to that of observing the opposite value,
$-W_{\tau}$, is
\begin{eqnarray}
  \frac{\Pi_{\tau}( W_{\tau} ) }  {\Pi_{\tau} (-W_{\tau})} & = & {\rm e}^{W_{\tau}} .
  \label{FT}
\end{eqnarray}   
Nevertheless, the class of systems obeying Eq.~(\ref{FT}) is unknown
-- even for stochastic dynamics, where FT is most easily
derived~\cite{Ku,LeSp} -- for it is not clear a priori when the
asymptotic large-deviation/long-time regime understood in Eq.~(\ref{FT})
is attainable, and whether it does generally reflect most situations
of physical interest.  In fact, genuine deviations from Eq.~(\ref{FT})
have been early observed~\cite{Se_1998} and are now well established
in various
contexts~\cite{ZaRuAn,ZaBoCuKu,Puglisi_1,Puglisi_2,Rakos,ViPuVu,Klages}.
Their general characterization -- if they are accidental in nature or
rather bring relevant informations about the stationary measure --
however, remains a widely open problem.

In this paper, I show that deviations respecting the
time reversal invariance of Eq.~(\ref{FT}), $W_{\tau} \to -W_{\tau}$,
generally occur in a large class of stochastic dynamics, and are a
signature of glassy correlations.
Their origin is traced back to the presence of heterogeneous anomalous
diffusion regimes that extend possibly beyond any range of physically
accessible values of $\tau$.  Two distinct small entropy production
limits will be considered: (i) vanishing forcing, i.e., near
equilibrium, and (ii) far from equilibrium when currents become small
at increasing drive, i.e. a negative resistance regime.  Deviations in
the former limit are induced by a transient subdiffusion, and tend to
decrease at long times. On the contrary, deviations in the latter
limit increase with $\tau$ and are due to a long-lived superdiffusion
regime.  Against our naive expectation based on the behavior of
relaxational glassy systems (where dynamics is typically sub-diffusive
in the aging regime), we find that mean-square current fluctuations
grow super-diffusively at high-density, even though the average
current becomes vanishingly small at increasing field.  This
particularly surprising behaviour is suggested to occur generally in
driven systems dominated by steric hindrance and cage effect. Systems
of experimental relevance include shear-thickening/jamming fluids,
where the glassy correlations are intimately related to dynamic
heterogeneity.
The existence of non-monotonic responses is a peculiar feature of
nonequilibrium steady state (NESS) and gives us the possibility of
testing FT beyond the linear-response, and in situations which are of
physical interest for driven soft matter.

The models we study are emerging as a new paradigm for the
interpretation of glassy phenomena~\cite{RiSo}. When driven into a
NESS they show a nonmonotonic dependence of the relaxation time on the
applied force, a feature which is qualitatively similar to the
behaviour of viscosity in sheared concentrated
suspensions~\cite{Se_2008}.
Two ingredients characterize the dynamics: (i) the cage effect--a
universal feature of glassy systems, which is implemented on a
coarse-grained scale through a local kinetic constraint, and (ii) a
nonconservative force which consists of a uniform and constant drive,
allowing for nonzero net current in the NESS. For simplicity, we
consider two-dimensional (2D) square lattice systems in which the
force is applied along a lattice axes, and the kinetic constraint
takes the following form: a randomly chosen particle can move to a
randomly chosen nearest neighbour site if i) the site is empty and ii)
the particle has at most two nearby particles before and after the
move. Further, if the move of a mobile particle is attempted in
direction opposite to the field, the probability is ${\rm e}^{-\beta
  E}$, otherwise it is $1$.

\begin{figure*}
\begin{center}
\includegraphics[width=5.8cm]{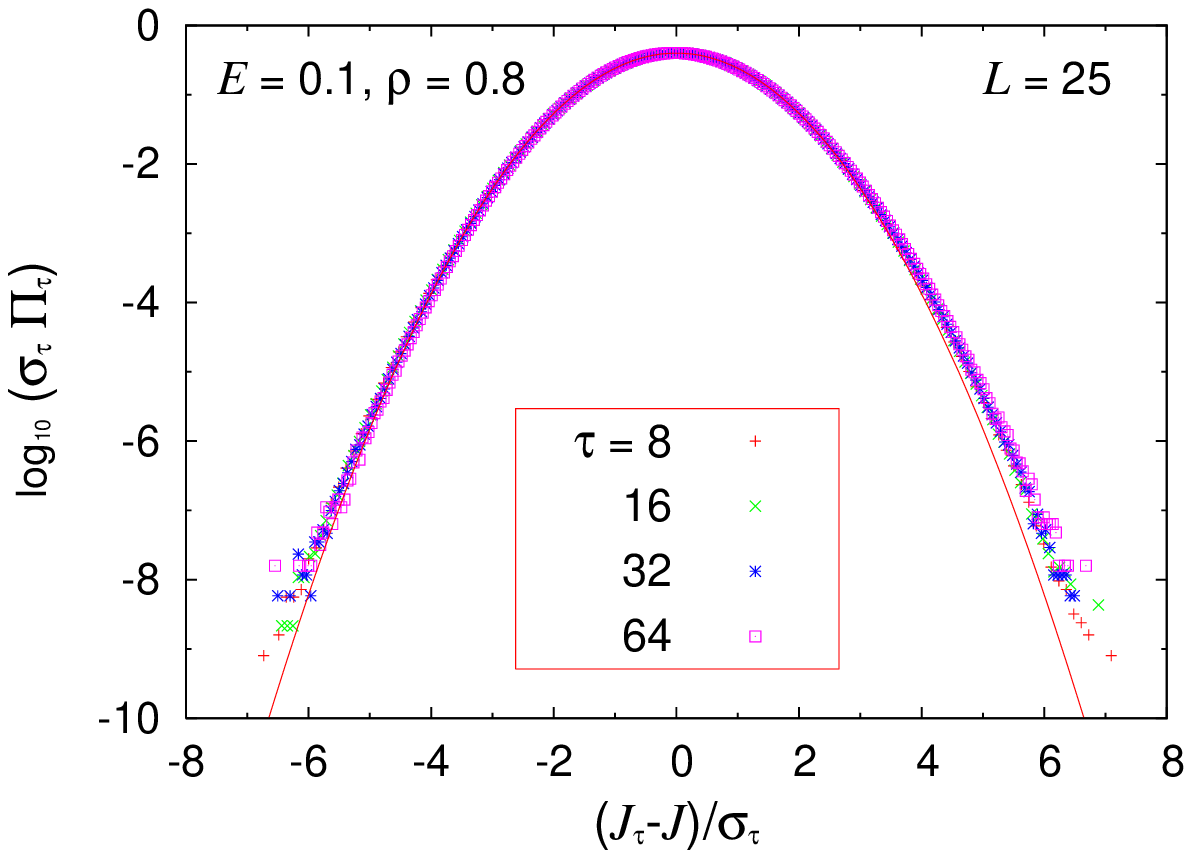}
\includegraphics[width=5.8cm]{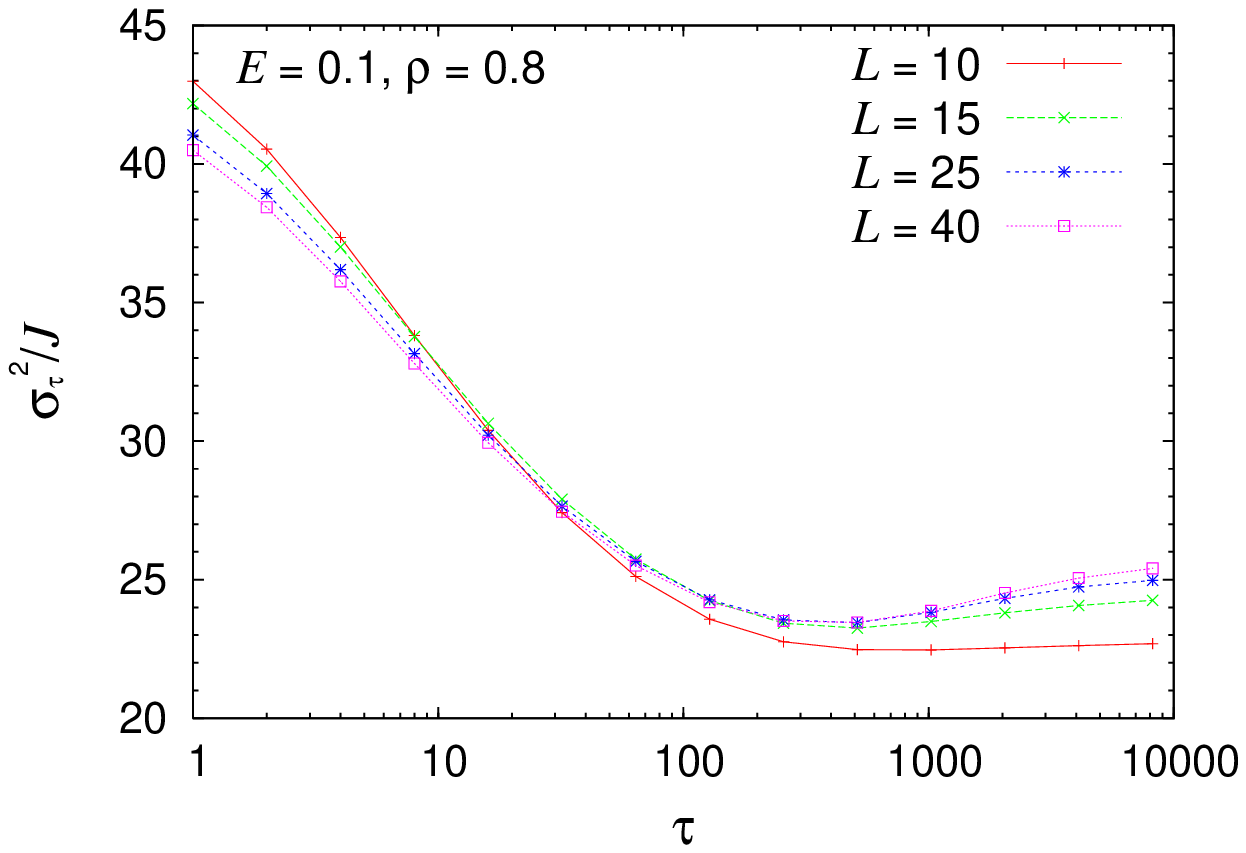}
\includegraphics[width=5.8cm]{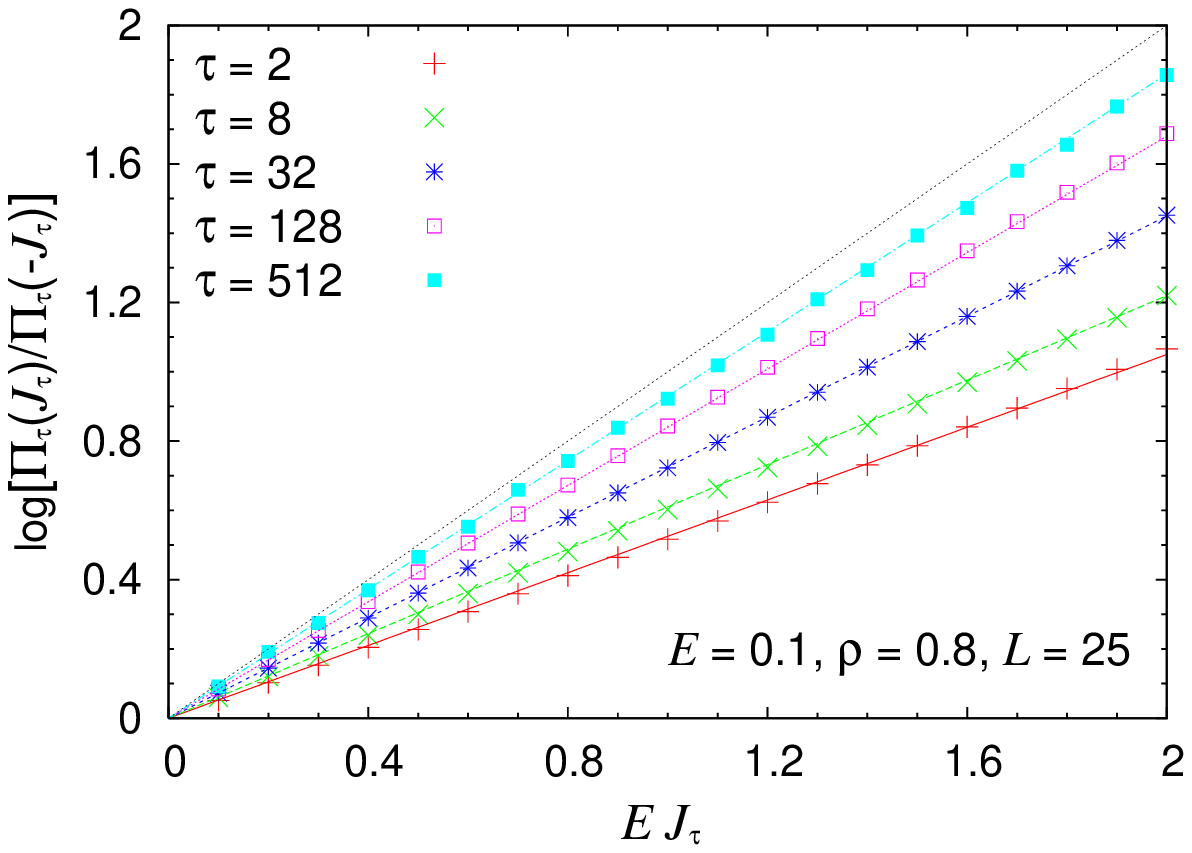}
\end{center}
\caption{2D constrained exclusion process driven by a small applied
  field $E=0.1$, at particle density $\rho=0.8$. The time $\tau$ is
  measured in MCs and $L$ is the linear system-size.  {\it Left}: PDF
  of current fluctuations in normalized unit.  Full line
  represents the Gaussian.  {\it Middle}: Temporal evolution of
  relative fluctuations for several system sizes. {\it Right}:
  Logarithmic probability ratio $\log(\Pi_{\tau}(J)/\Pi_{\tau}(-J))$
  vs. entropy production $W_{\tau}=E J_{\tau}$ ($\beta=1$). Full line, with
  slope $1$, represents the prediction of Eq.~(\ref{FT}).}
\label{E0.1}
\end{figure*}
The interplay of the two above ingredients gives a rather rich dynamic
behaviour~\cite{Se_2008}. At small forces, the cage around a given
particle is just slightly distorted allowing easily for particle
flow. At increasing forces, however, particles become generally more
caged by their neighbors, as moves against the field direction are
much less probable. Local rearrangements are thus more difficult and
transport more obstructed.  As we shall see, this situation leads to
non-Gaussian current fluctuations, growing heterogeneous
spatio-temporal correlations and anomalous diffusion regimes.

For locally reversible and irreducible Markov chains, such as the one
introduced above, the proof of FT is straightforward~\cite{LeSp}. One
considers the action
\begin{eqnarray} W_{\tau}(\{\sigma\}) = \sum_{t=0}^{\tau-1} \log
  \frac{w(\sigma_t,\sigma_{t+1})}{w(\sigma_{t+1},\sigma_t)},
\label{W}
\end{eqnarray} 
where $w(\sigma,\sigma') \ge 0$ are the transition probabilities for
jumping from configuration $\sigma$ to $\sigma'$. If the ``border
term'' $B = \log \left[ \mus(\sigma_{\tau})/\mus(\sigma_0) \right] $,
(where $\mus$ is the stationary measure), is subextensively small in
$\tau$, then the generating function of the action verifies
$\left\langle {\rm e}^{-\lambda W_{\tau}} \right\rangle = \left\langle
  {\rm e}^{-(1-\lambda) W_{\tau}} \right\rangle$, which is just
another form of FT.  In particular, when $\mus$ is flat over fixed
density configurations, then $B=0$ and FT holds at any time $\tau$.
This special case occurs in the asymmetric simple exclusion process
(ASEP) to which our model reduces in the absence of constraints. In
the presence of constraints, however the NESS measure is not trivial:
while the density profile is flat, particles are statistically more
clustered in the transverse direction at larger field. This
nonequilibrium fluctuation-induced attraction appears in the
transverse pair-correlation function and is a consequence of the more
hindered longitudinal transport at increasing field.  Nonetheless,
while the transverse diffusion slows down (due to the stronger
attraction), longitudinal diffusion is enhanced and becomes
superdiffusive over a growing range of times, at increasing field.
Neglecting border terms in this situation is not generally allowed
even for large $\tau$, and that is the ultimate reason of the observed
deviations from Eq.~(\ref{FT}). The importance of border terms is
discussed thoroughly in Ref.~\cite{Puglisi_2}. We now explore their
physical implications for driven systems with glassy dynamics.

\paragraph*{Fluctuation relation.--} 
To test Eq.~(\ref{FT}) we perform Monte Carlo simulations of the above
driven stochastic dynamics.  A system of linear size $L$ is
initialized with a uniform distribution of particles at fixed particle
density $\rho$, and is let to reach the NESS. Observables of interest
are evaluated over time intervals of duration $\tau$, along a
trajectory of motion lasting $10^7-10^9$ Monte Carlo sweeps (MCs),
depending on the system size. In particular, we consider the particle
current $J_{\tau}$, that is the signed number of jumps over $\tau$
along the applied field, and compute its probability density function
(PDF), $\Pi_{\tau}(J_{\tau})$.  Since $w(\sigma,\sigma') = \Theta({\rm
  constraint}) \, {\rm min} \{1,{\rm e}^{-\beta \vec{E} \cdot \vec{dr}
}\}$, local detailed balance holds irrespective of local kinetic
constraints. For mobile particles one has $
w(\sigma,\sigma')/w(\sigma',\sigma) = 1, {\rm e}^{\pm \beta E}$,
depending on the relative direction of unit displacement and applied
field ($\vec{E} \cdot \vec{dr} = 0, \pm E$). Whereas for immobile
particle, $w(\sigma,\sigma')/w(\sigma',\sigma) = 1$, no matter the
value of $\vec{E} \cdot \vec{dr}$. The action functional~(\ref{W}) can
then be easily identified as $W_{\tau}= \beta E J_{\tau}$,
consistently with the standard definition of entropy production
obtained from the time-dependent Gibbs entropy formula~\cite{LeSp}.
Thus, computing $\Pi_{\tau}(J_{\tau})$ allows for a direct check of
FT. Notice, that the absence of both potential and kinetic energy
terms makes the PDFs of current, work and heat exactly identical.

\begin{figure*}
\begin{center}
  \includegraphics[width=5.8cm]{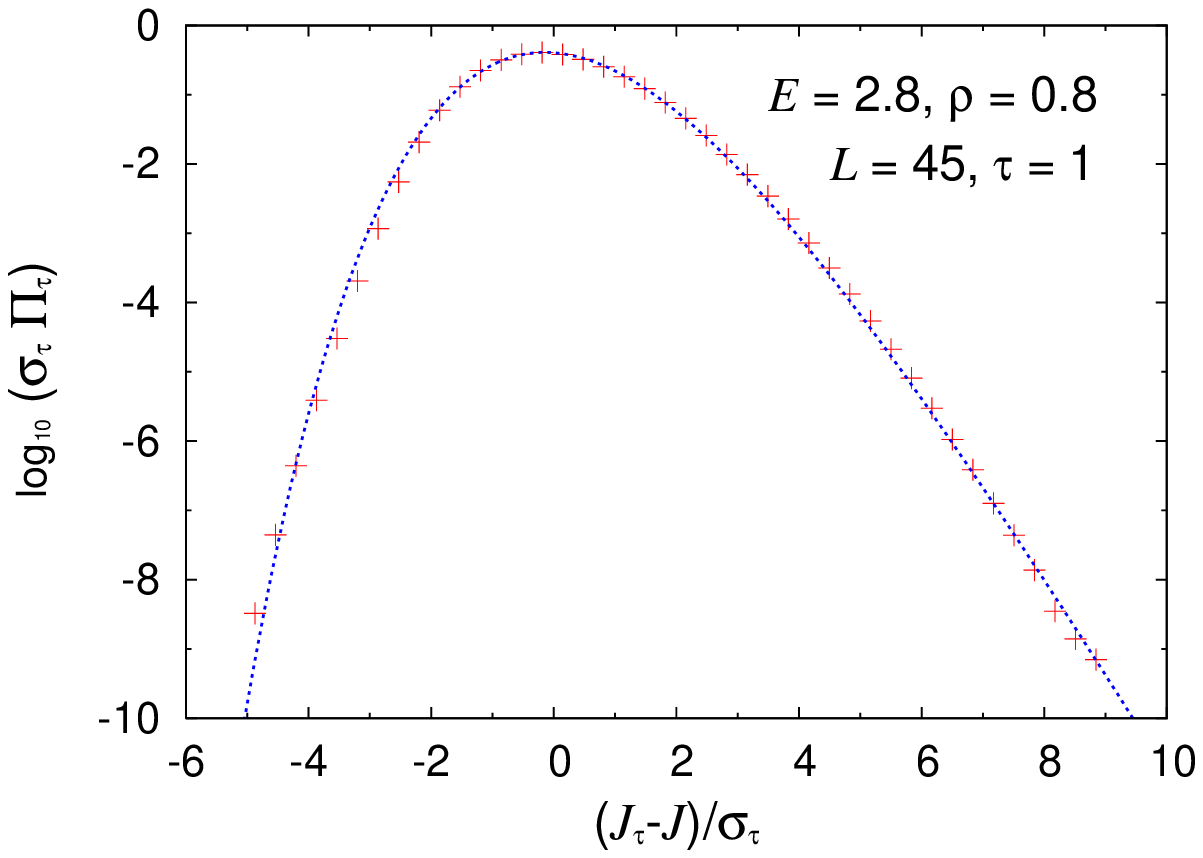}
  \includegraphics[width=5.8cm]{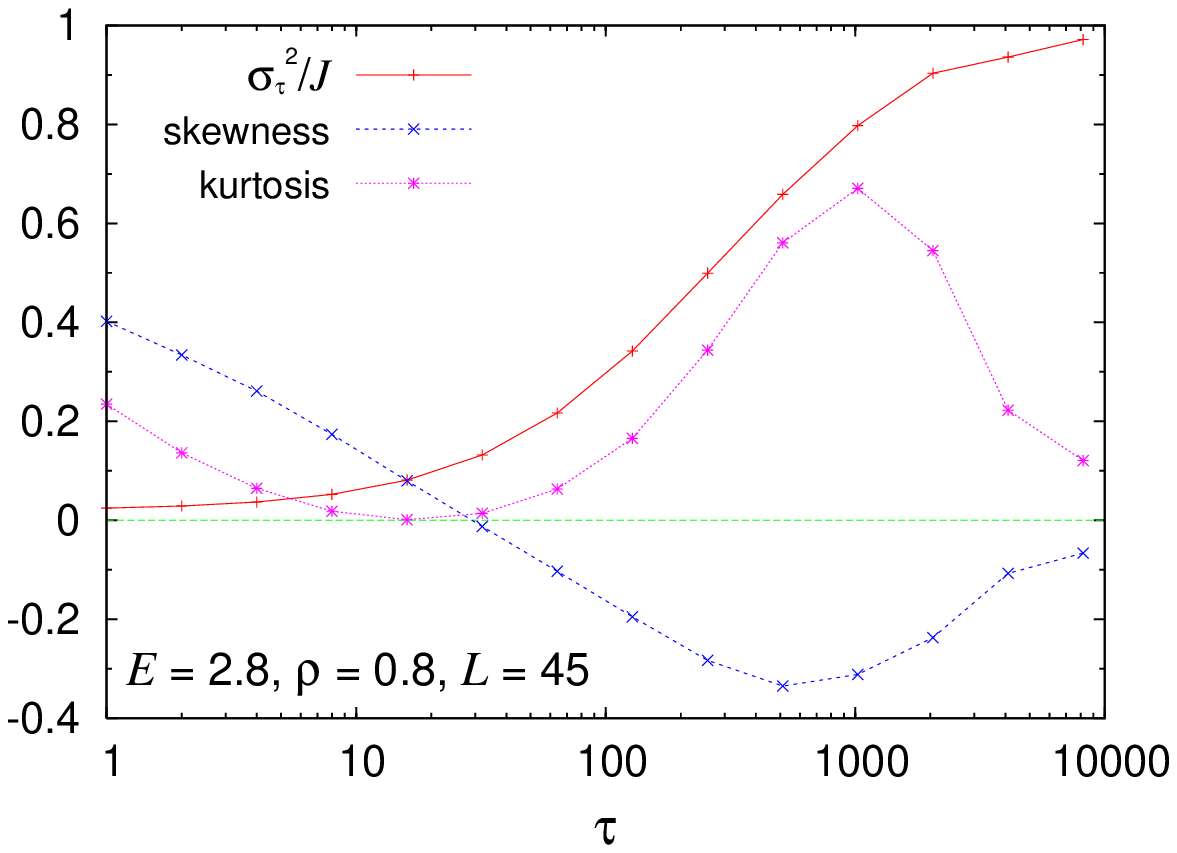}
  \includegraphics[width=5.8cm]{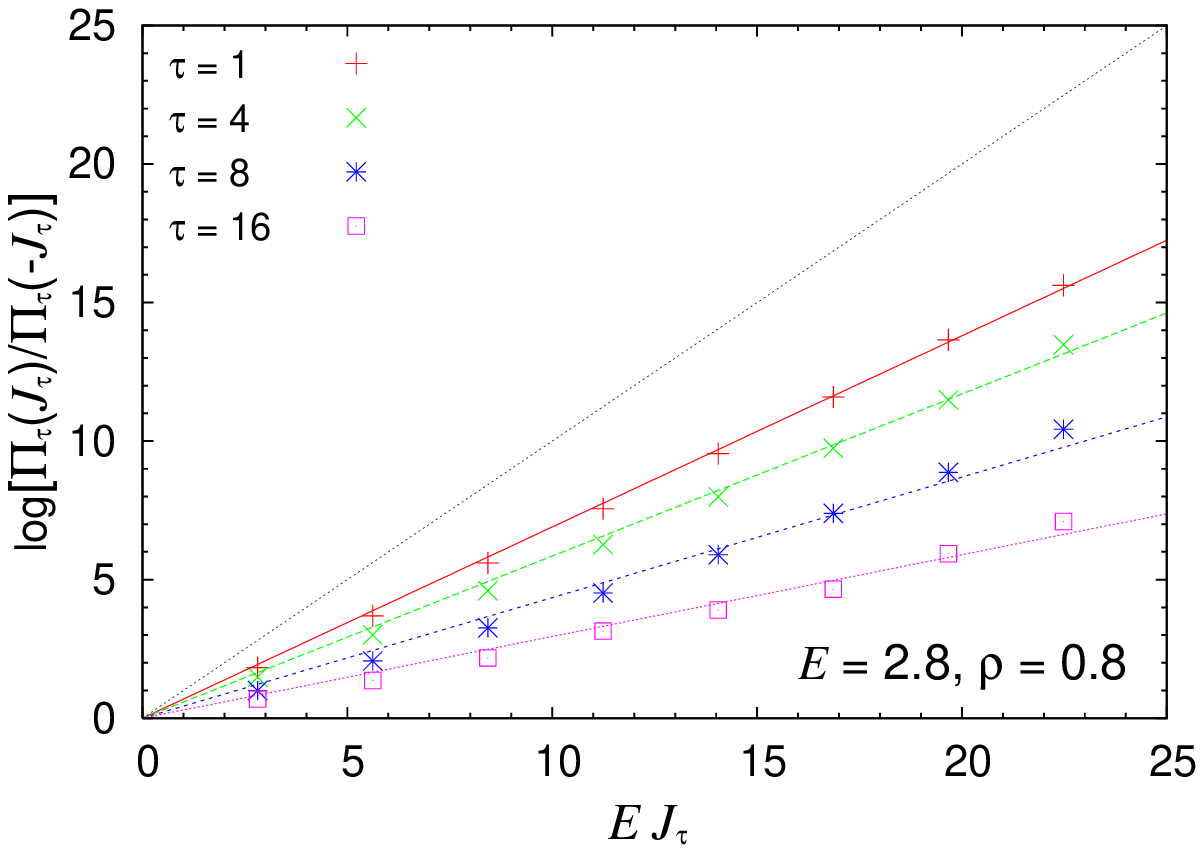}
\end{center}
\caption{Same as in Fig.~1, but in the negative resistance regime,
  with a large applied field $E=2.8$ and $L=45$.  {\it Left}: The full
  line is a generalised Gumbel function, with fitting parameter
  $a=11$, see~\cite{Portelli}.  {\it Middle}: Temporal evolution of
  PDF high-order moments. Notice that relative fluctuations are
  rescaled by a factor 65. For systems of linear size $L=45$, the
  time-scale for observing current-reversal events is at most of the
  order of $10^2$ MCs. {\it Right}: Deviations from Eq.~(\ref{FT})
  here increase with $\tau$.}
\label{E2.8}
\end{figure*}

We have first checked that in the standard ASEP current fluctuations
are Gaussian distributed and that Eq.~(\ref{FT}) is obeyed for time as
small as $\tau=1$ MCs, as expected because of flat measure.  Similar
behavior is found for constrained driven dynamics at small density. At
moderately higher densities and small fields, the particle motion
becomes weakly correlated and small non-Gaussian tails appear in the
PDF of current fluctuations, see~Fig.~\ref{E0.1} (left panel). Since
the average current $J=\langle J_{\tau} \rangle$ does not depend on
$\tau$, and the PDFs for various values of $\tau$ fall on the top of
each other when plotted in normalized units, one might naively expect
that the asymptotic large-deviation regime has been attained.
Actually, current fluctuations, $\sigma^2_{\tau}$, still retain a
dependence on $\tau$, and increasing the system size makes it harder
to reach the asymptotic value, Fig.~\ref{E0.1} (middle). The slow
decay of current fluctuations leads to deviations from Eq.~(\ref{FT})
which become smaller and smaller at increasing $\tau$,
see~Fig.~\ref{E0.1} (right).  Though such deviations are expected to
disappear at longer $\tau$, in fact recovering FT maybe difficult at
larger density, because the Ohmic regime shrinks~\cite{Se_2008}.

On approaching the negative resistance transport regime, which is a
simple rheological analog of shear-thickening behavior~\cite{Se_2008},
something more interesting takes place.  First, current fluctuations
becomes strongly non-Gaussian and asymmetrically distributed,
see~Fig.~\ref{E2.8} (left).  Similar asymmetric PDFs have been
observed in a wide range of systems~\cite{BeCl}.  Second, although the
average current quickly attains its asymptotic value, the PDF keeps
evolving with $\tau$, as shown by the behaviour of high-order moments
in~Fig.~\ref{E2.8} (middle).  Skewness and kurtosis generally display
a nonmonotonous dependence on $\tau$, which is suggestive of that
found in the local correlations during the aging dynamics of glassy
systems~\cite{Chamon}, and can be qualitatively understood as follows.
On short time scales most particles are blocked and only a small
fraction of them is able to move in the field direction, thus the PDF
is left-skewed. Local rearrangements involving backward steps,
however, are needed to sustain the system flow.  Indeed, on
intermediate time scales, current reversal events become more probable
and, correspondingly, the PDF is less asymmetric. On longer time
scales, most particles have moved and the probability of observing
negative current is pretty small (although current reversal events
keep happening on shorter time scales). The PDF thus changes to a
right-skewed shape before eventually reaching the Gaussian form
typical of diffusive, uncorrelated motion.
Figure~\ref{E2.8} (middle) shows that the latter diffusive regime occurs
at times that are much beyond the time-scale required for observing
negative currents (at least two decades larger in the most favourable
case, corresponding to parameters of Fig.~\ref{E2.8}).
Therefore, the asymptotic regime of large-deviation understood in
Eq.~(\ref{FT}) is hardly attained, even for such small systems.
What type of deviations should we expect in the present case and what
features of NESS they possibly encode?  In spite of PDF asymmetry, we
find that deviations from Eq.~(\ref{FT}) are still well described by
straight lines with a slope smaller than one, and their temporal
evolution is mainly determined by current fluctuations,
$\sigma^2_{\tau}$.  However, unlike the previously discussed linear
transport regime, the slope appears to {\em decrease} at longer
$\tau$, see Fig.~\ref{E2.8} (right).
The numerical extrapolation of the asymptotic slope is difficult in
this regime, because the accessible values of $\tau$ are limited by
the larger system-size used.
In fact, sampling the PDF requires some care at high-density and
large-force: to avoid finite-size induced dynamical blocking effects
(leading to bimodal PDFs) one needs to use moderately larger
systems. Since, this obviously decreases the probability of observing
current reversal, we consider density barely larger than the negative
resistance threshold.  Even in this most favorable situation (in which
the saturation current is finite~\cite{Se_2008}), current reversal
events become unobservable well before the asymptotic regime is
attained.  The trend of high-order moments thus makes us confident
that Eq.~(\ref{FT}) will never be recovered in any realistic
simulation.

\begin{figure}
\begin{center}
  \includegraphics[width=6.8cm]{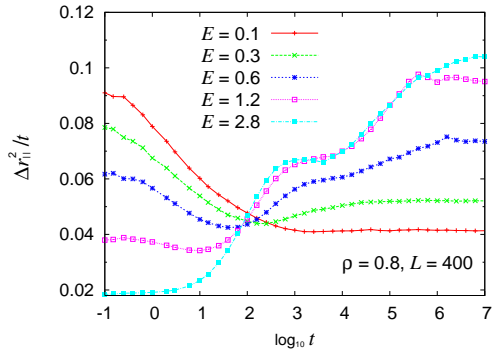}
\end{center}
\caption{Time averaged longitudinal mean-square particle displacement,
  $\Delta r_{||}^2/t$, vs. time $t$.}
\label{Fig_msd}
\end{figure}

\paragraph*{Heterogeneous anomalous diffusion.--} 
Further insight into the nature of deviations from Eq.~(\ref{FT}) is
obtained by looking at the behavior of the longitudinal mean square
displacement relative to the center of mass, $\Delta r^2_{||}(t)$.
Since the latter quantity is proportional to $\sigma^2_{\tau}$, the
physical origin of the above behaviour can be traced back to the
existence of heterogeneous anomalous diffusion regimes, see
Fig.~\ref{Fig_msd}.  In the linear transport regime, $\Delta
r^2_{||}(t)/t$ first decreases with $t$, and then reaches an
asymptotic diffusive plateau, that increases with the applied field.
This enhanced diffusion behaviour is also displayed by a driven
particle probing an equilibrium glassy environment~\cite{Jack}.
On approaching the negative resistance regime, the early subdiffusion
range tends to shrink and connects to the late normal (enhanced)
diffusion through an intermediate (logarithmic) superdiffusive
transient. Similar nonmonotonic changes of the effective anomalous
diffusion exponent have been observed in simulations of biased
Brownian motion in a crowded environment~\cite{Stauffer}, and are
consistent with some continuous-time random-walk models~\cite{Hughes}.
In the negative resistance regime, subdiffusion is strongly
suppressed, and one only observes a long-lived superdiffusion behavior,
see Fig.~\ref{Fig_msd}.  Although, we expect normal diffusion to occur
eventually for any finite system,
interestingly, we find that the crossover time to normal diffusion
increases with the system size.  The appearance of superdiffusion is
particularly intriguing in this context. It signals the simultaneous
onset of long-time correlations in the motion of particles and the
growth of their spatial correlations, as shown by the increasing peak
of dynamical susceptibility~\cite{Se_2008}.  A strikingly similar
connection between anomalous diffusion and dynamic heterogeneity has
been recently observed in vibrated granular packings, albeit in a
distinct jamming/rigidity regime~\cite{LeDaBiBo}, see
also~\cite{Head}.  It is tempting to conjecture, that the most
peculiar feature distinguishing the approach to the
near-jamming/shear-thickening/negative-resistance regime from the glassy
behavior, is that in the latter diffusivity tends to vanish, while in
the former it grows unboundedly.

Finally, it should be remarked that the results reported above are not
specific of 2D: they were observed in 3D, for different direction of
the applied force/lattice geometry, and for different types of kinetic
constraints as well.

\medskip

To conclude, our results generally show that for a large class of
driven stochastic dynamics, the asymptotic regime in which the
steady-state FT holds, exceeds any reasonable physical time-scale,
even for systems of modest size.  The presence of heterogeneous
superdiffusion in the negative resistance transport regime possibly
suggests that the large-deviation function might not even exist in the
usual sense, if the large-size limit is taken before the long-time
one~\cite{Gar}. In spite of the above limitations, we find that
deviations from (\ref{FT}) have several interesting features.  First,
they encode important physical properties, such as heterogeneous
dynamics and anomalous diffusion.  The strong ``finite-time''
deviations from (\ref{FT}) are a signature of growing long-range
dynamical correlations both in time and space, and are arguably a
universal feature inherent to driven systems with glassy dynamics,
such as shear-thickening/jamming fluids.  Second, their linear form is
the simplest one respecting the time-reversal invariance of
(\ref{FT}), though the asymptotic limit understood in FT is not
representative of the physical situation. Finally, they are highly
suggestive of the notion of correlation-scale dependent effective
temperature~\cite{CuKuPe}. While the occurrence of longitudinal
superdiffusion seems to prevent the applicability of such an appealing
concept, we find that transverse fluctuation dynamics is well
described by a generalised Einstein relation (in some analogy to
systems of driven vortices with random pinning~\cite{Kolton}). Whether
a modified form of Einstein relation~\cite{Evans,Speck} does hold for
longitudinal fluctuations remains to be seen.

\bigskip

\noindent M.S. thanks G. Biroli, L.F. Cugliandolo, J. Kurchan and
F. Zamponi for discussions, and acknowledges the support of a ``Chaire
Michelin''.


\begin{thebibliography}{35}

\bibitem{HaSc} R.J. Harris and G. Sch\"utz, J. Stat. Mech.: Theory
  Exp.  P07020 (2007)

\bibitem{Ri} F. Ritort, Adv. Chem. Phys. {\bf 137}, 31 (2008)

\bibitem{Vulpio} U. Marini Bettolo Marconi, A. Puglisi, L. Rondoni and
  A. Vulpiani, Phys. Rep. {\bf 461}, 111 (2008)
  
\bibitem{Ku} J. Kurchan, J. Phys. A {\bf 31}, 3719 (1998)
  
\bibitem{LeSp} J.L. Lebowitz and H. Spohn, J. Stat. Phys. {\bf 95},
  333 (1999)

\bibitem{Se_1998} M. Sellitto, {\tt arXiv:cond-mat/9809186}
  
\bibitem{ZaRuAn} F. Zamponi, G. Ruocco and L. Angelani, Phys. Rev. E
  {\bf 71}, 020101(R) (2005)
 
\bibitem{ZaBoCuKu} F. Zamponi, F. Bonetto, L.F.  Cugliandolo and
  J. Kurchan, J. Stat. Mech.: Theory Exp. (2005) P09013 

\bibitem{Puglisi_1} A. Puglisi, P. Visco, E. Trizac and F. van
  Wijland, Phys. Rev. E {\bf 73}, 021301 (2006)

\bibitem{Puglisi_2} A. Puglisi, L. Rondoni and Vulpiani,
  J. Stat. Mech.: Theory Exp. (2006) P08010 

\bibitem{Rakos} A. R\'akos and R.J. Harris, J. Stat. Mech.: Theory
  Exp. (2008) P05005 

\bibitem{ViPuVu} D. Villamaina, A. Puglisi and A. Vulpiani,
  J. Stat. Mech.: Theory Exp. (2008) L10001 

\bibitem{Klages} A.V. Chechkin and R. Klages, J. Stat. Mech.: Theory
  Exp. (2009) L03002 

\bibitem{RiSo} F. Ritort and P. Sollich, Adv. Phys. {\bf 52}, 219
  (2003)

\bibitem{Se_2008} M. Sellitto, Phys. Rev. Lett. {\bf 101}, 048301
  (2008)

\bibitem{Chamon} C. Chamon et al.,
  J. Chem. Phys. {\bf 121}, 10120 (2004)

\bibitem{Portelli} B. Portelli et al., 
  Phys. Rev. E {\bf 64}, 036111 (2001)

\bibitem{BeCl} M. Clusel and E. Bertin, Int. J. Mod. Phys. B {\bf 22},
  3311 (2008)

\bibitem{Jack} R.L. Jack, D. Kelsey, J.P. Garrahan and D. Chandler,
  Phys. Rev. E {\bf 78}, 011506 (2008)

\bibitem{Stauffer} D. Stauffer, C. Schulze and D.W. Hermann,
  J. Biol. Phys. {\bf 33}, 305 (2007)

\bibitem{Hughes} B. Hughes, {\em Random Walks and Random Environments}
  (Clarendon Press, Oxford, 1995)

\bibitem{LeDaBiBo} F. Lechenault, O. Dauchot, G. Biroli and
  J.-P. Bouchaud, Europhys. Lett. {\bf 83}, 46003 (2008)

\bibitem{Head} D. Head, Phys. Rev. Lett. {\bf 102}, 138001 (2009)

\bibitem{Gar} In this respect, it would be interesting to extend to
  driven systems the Ruelle thermodynamic formalism, as done in:
  J.P. Garrahan, R.L. Jack, V. Lecomte, E. Pitard, K. van Duijvendijk,
  and F. van Wijland, Phys. Rev. Lett. {\bf 98}, 195702 (2007)

\bibitem{CuKuPe} L.F. Cugliandolo, J. Kurchan and L. Peliti, Phys.
  Rev. E {\bf 55}, 3898 (1997)
  
\bibitem{Kolton} A.B. Kolton, R. Exartier, L.F. Cugliandolo,
  D. Dom\'inguez, and N. Gronbech-Jensen, Phys. Rev. Lett. {\bf 89},
  227001 (2002)

\bibitem{Evans} T. Hanney and M. Evans, J. Stat. Phys. {\bf 111}, 1377
  (2003)

\bibitem{Speck} T. Speck and U. Seifert, Europhys. Lett. {\bf 74}, 391 (2006)

\end{thebibliography}
\end{document}